\documentclass[showpacs,amsmath,amssymb,aps,prd, preprint,groupedaddress,superscriptaddress,nofootinbib]{revtex4-1} 
\usepackage[latin1]{inputenc}
\usepackage{graphicx}
\usepackage{tensor}
\usepackage{bm}

\newcommand{\e}[1]{\tensor{e}{#1}}
\newcommand{\eb}{{\bf e}}
\newcommand{\dual}{\bm{\theta}}
\newcommand{\h}[1]{\tensor{h}{#1}}
\newcommand{\torsion}[1]{\tensor{T}{#1} }
\newcommand{\superpotential}[1]{\tensor{\Sigma}{#1} }
\newcommand{\nablab}{\bm{\nabla}}
\newcommand{\rcd}{\mathring{\nabla}}
\newcommand{\rcdb}{\mathring{\bm{\nabla}}}
\newcommand{\p}[1]{\tensor{\partial}{#1}}
\newcommand{\R}[1]{\tensor{\mathring{R}}{#1}}
\newcommand{\D}{\mathring{D}}
\newcommand{\chr}[1]{\tensor{\mathring{\Gamma}}{#1}}
\newcommand{\sconnection}[1]{\tensor{\omega}{#1}}
\newcommand{\rbox}{\mathring{\Box}}

\begin{document}


\title{The energy-momentum tensor of gravitational waves, Wyman spacetime and freely falling observers }


\author{J. B. Formiga}
\email[]{jansen@fisica.ufpb.br}
\affiliation{Departamento de Física, Universidade Federal da Paraíba, Caixa Postal 5008, 58051-970 João Pessoa, Pb, Brazil}

\date{\today}


\date{\today}

\begin{abstract}
A good definition for the energy momentum tensor of gravity (EMTG) in General Relativity (GR) is a hard, if not impossible, task. On the other hand, in its teleparallel version, known as The Teleparallel Equivalent of General Relativity (TEGR), one can define the EMTG in a very satisfactory way. In this paper, it is proved that the EMTG of TEGR  for linearized gravitational waves (GWs) is the same as the version of GR that is usually given in the literature. In addition, the exact version of the EMTG for a $pp-$wave with a $+$ polarization is obtained in a freely falling frame (FFF). Unlike the previous case, the energy density can be either positive or negative, depending on the details of the wave. The gravitational energy density for the Wyman spacetimes is obtained both in a static frame and in a FFF. It turns out that observers in free fall can measure the effects of gravity.
\end{abstract}


\maketitle

\section{Introduction}
Energy is one of the most important concept in physics, since it takes many forms and is involved in all physical processes. It does not matter whether we are dealing with a classical or a quantum system, we just have to deal with it. Nonetheless, the definition and localization of the energy of some fields may become a problem for physicists. This is exactly what happens with the energy and the momentum of a gravitational field. There have been many controversies over this issue in General Relativity (GR) \cite{0264-9381-9-7-009,Wald:1984rg,Cooperstock2018} and it seems highly unlikely that, in the standard metrical formulation of GR, one will be able to make a satisfactory definition of the EMTG \cite{ANDP:ANDP201200272,doi:10.1002/andp.201700175}. To overcome this difficulty, one has to use the tetrad formalism. This is best done in the framework of teleparallelism, where the fundamental quantity is the tetrad field.

Despite the controversy over the definition of an EMTG in GR, the definition of such a tensor for GWs are very common in the literature \cite{Gravitation,MicheleMaggiore1,creighton2012gravitational}. In general, what is done is to split the metric tensor $g_{\mu\nu}$ into a background  $\bar{g}_{\mu\nu}$ and in a perturbation $h_{\mu\nu}$ which represents ripples in the spacetime; then one sees how $h_{\mu\nu}$ contributes to the curvature of the background. Of course, this is not always possible and, therefore, the definition is not general (it does not include the cases where the spacetime satisfies a wave equation, but the conditions to be interpreted as GW are not valid\footnote{For those conditions, see section 35.7 of Ref.~\cite{Gravitation}.}) \cite{Gravitation}. Furthermore, as far as the author is aware, there is no exact expression for the energy of a GW in GR.

The most famous theory based on the concept of distant parallelism  is known as The Teleparallel Equivalent of General (TEGR). Since this theory is very similar to GR and allows for a reasonable definition of the EMTG, one might ask whether we can make a parallel between the EMTG of the TEGR and the results that are expected from GR. In Ref.~\cite{ANDP:ANDP201200272}, and references therein, the authors obtain the ADM energy for asymptotic spacetimes and show that, in a FFF, the EMTG of the Schwarzschild spacetime vanishes, which is in accordance with the equivalence principle. In Ref.~\cite{universe4070074} they study the variation of the energy of free particles under a $pp$-wave spacetime and, in Ref.~\cite{PhysRevD.78.047502}, they present the exact form of the energy of the wave as measured by a set of static observers. They conclude that the energy is always negative; apparently, no parallel between this energy and that of GR can be made. Nonetheless, in Ref.~\cite{0264-9381-20-21-008}, the average flux densities per unit period, flowing along some directions, are calculated and the results match those of GR. Now, it comes the question: is the EMTG for GWs in TEGR the same as the one of GR? The answer to this question is ``yes''.

The main goal of this paper is to show that the linearized energy-momentum tensor of the TEGR agrees with the common definition of its counterpart in GR, Sec.~\ref{06072018a}. It turns out that this energy is positive, as one would expert. In Sec.~\ref{06072018b}, the EMTG for a general $pp$-wave spacetime with a $+$ polarization is obtained. The result shows that the energy can be positive or negative, depending on the form of the wave. An interesting feature of this energy is that the observers who measure it are in free fall, which goes against our intuitive idea that this kind of observers cannot measure the effects of gravity. To make sure that this is not a feature of nonstatic spacetimes, it is shown in Sec.~\ref{30072018a} that the Janis-Newman-Winicour-Wyman  (JNW) spacetime also possesses this property. The energy for static observers is also calculated in Sec.~\ref{30072018b}.  A brief review of teleparallel theories is given in the next section.  

The notation used throughout this paper is basically the same as that of Ref.~\cite{ANDP:ANDP201200272}: $\e{_a^\mu}$ represents the components of the frame $\eb_a$ in the coordinate basis $\p{_\mu}$, while $\e{^a_\mu}$ are the components of the dual basis $\dual^a$, written in terms of $dx^{\mu}$; Latin indices run from $(0)$ to $(3)$, except for letters in the middle of the alphabet, which run over spatial components only; as usual, Greek letters run from $0$ to $3$. Unlike Ref.~\cite{ANDP:ANDP201200272}, the signature of the metric is $(+1,-1,-1,-1)$.

\section{Teleparallel Gravity}\label{30072018c}
Teleparallel theories are based upon the idea of absolute parallelism, which means that the curvature tensor vanishes. In general, one assumes that there exists a frame $\e{_a^\mu}$ satisfying the condition
\begin{equation}
\nabla_{\lambda}\e{_a^\mu}=0, \label{22072018a}
\end{equation}
where $\nablab$ is the affine connection and $\nabla_{\lambda}\e{_a^\mu}$ are the components of $\nablab_{\lambda}\eb_a$. When written in a coordinate basis, this connection takes the form $\tensor{\Gamma}{^\lambda_\mu_\nu}=\e{^a^\lambda}\partial_{\mu}\e{_a_\nu}$. An affine connection satisfying Eq.~(\ref{22072018a}) is known as the Weitzenböck connection. When written in terms of $\eb_a$, it vanishes: $\sconnection{^a_b_c}:=\left<\dual^a,\nablab_{b}\eb_c\right>=\e{_b^\lambda}\left<\dual^a,\nablab_{\lambda}\eb_c\right>=\e{_b^\lambda}\left<\dual^a,\left(\nabla_{\lambda}\e{_c^\mu}\right)\p{_\mu}\right>=0$. On the other hand, in terms of another basis $\tilde{\eb}_a$ that is related to $\eb_a$ by a local Lorentz transformation, $\tilde{\eb}_c=\tensor{\Lambda}{^d_c}\eb_d$, this connection does not vanish: $\tensor{\tilde{\omega}}{^a_b_c}=\left<\tilde{\dual}^a,\nablab_{\tilde{\eb}_b}\tilde{\eb}_c\right>=\tensor{\Lambda}{_d^a}\tensor{\tilde{e}}{_b^\lambda}\left<\dual^d,\left(\nablab_{\lambda}\tensor{\Lambda}{^e_c}\right)\eb_e \right>=\tensor{\tilde{e}}{_b^\mu}\tensor{\Lambda}{_d^a}\p{_\mu}\tensor{\Lambda}{^d_c}$. 
This is the reason why this connection is sometimes called the inertial spin connection \cite{0264-9381-34-14-145013}.

It follows from Eq.~(\ref{22072018a}) that the nonmetricity tensor must vanish as well. So, to have a nontrivial geometry, we must assume that torsion describes gravity. In terms of $\eb_a$, the torsion components become
\begin{equation}
\tensor{T}{^a_b_c}=2\tensor{e}{_b^\mu}\tensor{e}{_c^\nu}   \p{_{[\mu|}}\e{^a_{|\nu]}}, \label{22072018b}
\end{equation}
where $\p{_{[\mu|}}\e{^a_{|\nu]}}=(1/2)(\p{_\mu}\e{^a_\nu}-\p{_\nu}\e{^a_\mu})$.

The TEGR corresponds to the theory that uses the identity $\R{}=-T+2\rcd{_\mu}T^{\mu}$, where $\R{}$ is the Ricci scalar of the Riemannian connection $\rcd$ (the one of General Relativity) and $T$ is the torsion scalar\footnote{As in Ref.~\cite{ANDP:ANDP201200272}, the conventions used here are $T_\mu=\tensor{T}{^b_b_\mu}$ and $T=\frac{1}{4}T^{abc}T_{abc}+\frac{1}{2}T^{abc}T_{bac}-T^aT_a$.}, to obtain field equations that are exactly the Einstein field equations written in terms of the field $\e{_a^\mu}$. Therefore, the TEGR has the same solutions as GR, regardless of the tetrad field we choose.

In principle, any set of tetrad can be chosen to satisfy Eq~(\ref{22072018a}). However, some of them are problematic in the sense that the torsion tensor does not vanish in the absence of gravity. This happens because the expression (\ref{22072018b}) is measuring the anholonomicity of $\e{_a^\mu}$. So, if the parallel frame is anholonomic even in the absence of gravity, then the torsion tensor will not vanish. This problem is overcome by avoiding these type of frames.

An interesting aspect of the TEGR is the definition of an EMTG \cite{ANDP:ANDP201200272}. This tensor is defined as
\begin{equation}
t^{\lambda\mu}=kc\left(4\Sigma^{bc\lambda}\tensor{T}{_b_c^\mu}-g^{\lambda\mu}\Sigma^{bcd}T_{bcd} \right), \label{22072018c}
\end{equation}
where
\begin{equation}
\Sigma^{abc}=\frac{1}{4}\left( T^{abc}+T^{bac}-T^{cab}\right)+\frac{1}{2}\left(\eta^{ac}T^b-\eta^{ab}T^c \right) \label{22072018d}
\end{equation}
is the superpotential and $k=c^3/(16\pi G)$ ($G$ is the gravitational constant and $c$ the speed of light). Note that $t^{\lambda\mu}$ is tracefree.

The field equations of the theory can be written in the form
\begin{equation}
\p{_\nu}\left(e\Sigma^{a\lambda\nu} \right)=\frac{1}{4kc}e\e{^a_\mu}\left(t^{\lambda\mu}+T^{\lambda\mu} \right) \label{22072018e}
\end{equation}
with $T^{\lambda\mu}$ being the energy-momentum  tensor of matter and  $e=\textrm{det}(\e{^a_\mu})$. Since $\Sigma^{abc}=-\Sigma^{acb}$, we have $\p{_\lambda}\p{_\nu}\left(e\Sigma^{a\lambda\nu} \right)=0$ $\Rightarrow$ $ \p{_\lambda}\left[e\e{^a_\mu}\left(t^{\lambda\mu}+T^{\lambda\mu} \right)\right]=0$. This is a true energy--momentum conservation equation, which is basically the reason why we interpret $t^{\lambda\mu}$ as the gravitational energy--momentum density. In addition, one might also interpret the expression
\begin{equation}
P^a=\int_V d^3x e\e{^a_\mu}\left(t^{0\mu}+T^{0\mu} \right) \label{22072018f}
\end{equation}
as representing the total energy--momentum within a three--dimensional volume $V$. In view of Gauss theorem, the left-hand side of Eq.~(\ref{22072018e}) can be used to rewrite $P^a$ in the form
\begin{equation}
P^a=4kc\oint_S dS_i e\Sigma^{a0i},\label{22072018g}
\end{equation}
where $i=1,2,3$ and $S$ is the boundary of $V$. The total energy of the system is identified with the component $P^{(0)}$: 
\begin{equation}
E=4kc\oint_S dS_i e\Sigma^{(0)0i}.\label{23072018a}
\end{equation}

The definitions (\ref{22072018f})-(\ref{23072018a}) depend on the frame we choose. Thus, the values of $P^a$ and $t^{\lambda a}$ will depend on the observers that are measuring the energy. This is so because the field $\e{_a^\mu}$ is implicitly associated with a set of observers. To be more precise, it is associated with the observers whose velocity along a curve $x^{\mu}(\tau)$, $u^{\mu}=dx^{\mu}/d\tau$, coincides with $c\e{_{(0)}^\mu}$ and carry a spatial frame $\e{_i^\mu}$ ($i=1,2,3$). To better understand these observers, one might use the antisymmetric acceleration tensor $\phi_{ab}$, defined through the relation
\begin{equation}
\frac{\D\e{_a^\mu}}{d\tau}=c\e{_{(0)}^\lambda}\rcd_{\lambda}\e{_a^\mu}=\tensor{\phi}{_a^b}\e{_b^\mu}, \label{23072018b}
\end{equation} 
where $\D\e{_a^\mu}/d\tau$ are the components of the Riemannian covariant derivative along $x^{\mu}(\tau)$, namely $\rcdb_{(d/d\tau)}\eb_a$. One says that a frame is in free fall when $\phi_{ab}=0$. Notice that this condition also implies that the frame is nonrotating.

\section{GW Energy Density in the TEGR}\label{26072018a}

\subsection{Linearized plane GWs }\label{06072018a}   
To show that\footnote{Notice that we use $t^{\lambda a}$ instead of $t^{\lambda\mu}$ because the conservation equation is $\p{_\lambda}(et^{\lambda a})=0$ (in vacuum). } $t^{\lambda a}$ yields the same result as that of GR, let us take our frame in the form
\begin{equation}
\e{_a_\mu}\approx \eta_{a\mu}+\frac{1}{2}\varepsilon\h{_a_\mu}, \label{22062018a}
\end{equation}
where $\eta_{a\mu}=diag(1,-1,-1,-,1)$ (Minkowski metric tensor),  $\varepsilon$ is a very small dimensionless parameter, and $\h{_a_\mu}$ are functions that account for the perturbations produced by the wave. It follows immediately from Eq.~(\ref{22062018a}) that
\begin{eqnarray}
g_{\mu\nu}\approx \eta_{\mu\nu}+\varepsilon\h{_\mu_\nu}&,\quad g^{\mu\nu}\approx \eta^{\mu\nu}-\varepsilon\h{^\mu^\nu}. \label{22062018b}
\end{eqnarray}
For the sake of simplicity, we can assume that the wave propagates along the z axis and use $u= t-z/c$. In this case, it is possible to find a Cartesian coordinate system where only the components $\h{_1_1}(u)$, $\h{_1_2}(u)=\h{_2_1}(u)$ and $\h{_2_2}(u)=-\h{_1_1}(u)$ are nonzero. A straightforward calculation shows that, neglecting terms of second order or higher in $\varepsilon$, the tetrad $\e{_a_\mu}$ becomes
\begin{equation}
\e{_a_\mu}\approx\left(
\begin{array}{cccc}
1 & 0 & 0 & 0\\
0 & -1+\frac{1}{2}\varepsilon\h{_1_1} & \frac{1}{2}\varepsilon\h{_1_2} & 0\\
0 & \frac{1}{2}\varepsilon\h{_1_2} & -1-\frac{1}{2}\varepsilon\h{_1_1} & 0\\
0 & 0 & 0 & -1
\end{array}\right). \label{22062018c}
\end{equation}
If we raise the Greek index with the metric $g^{\mu\nu}$, we will obtain
\begin{equation}
\e{_a^\mu}\approx\left(
\begin{array}{cccc}
1 & 0 & 0 & 0\\
0 & 1+\frac{1}{2}\varepsilon\h{_1_1} & \frac{1}{2}\varepsilon\h{_1_2} & 0\\
0 & \frac{1}{2}\varepsilon\h{_1_2} & 1-\frac{1}{2}\varepsilon\h{_1_1} & 0\\
0 & 0 & 0 & 1
\end{array}\right). \label{22062018d} 
\end{equation}
This frame is not only adapted to static observers but it is also a FFF to first order in $\varepsilon$.

The torsion components are
\begin{eqnarray}
\torsion{^{(1)}_{(1)(0)}}=\torsion{^{(1)}_{(3)(1)}}=\torsion{^{(2)}_{(0)(2)}}=\torsion{^{(2)}_{(2)(3)}}=(\varepsilon/2c)\dot{h}_{11}, \label{23062018a}
\\
\torsion{^{(2)}_{(1)(0)}}=\torsion{^{(1)}_{(2)(0)}}=\torsion{^{(2)}_{(3)(1)}}=\torsion{^{(1)}_{(3)(2)}}=(\varepsilon/2c) \dot{h}_{12}, \label{23062018b}
\end{eqnarray}
where the dot denotes $\p{_t}$. Since these components are written in a tetrad basis, we can easily raise them with $\eta^{ab}$ and use Eq.~(\ref{22072018d}) to obtain
\begin{eqnarray}
\superpotential{^{(1)(0)(1)}}=\superpotential{^{(1)(3)(1)}}=\superpotential{^{(2)(2)(0)}}=\superpotential{^{(2)(2)(3)}}=(\varepsilon/4c)\dot{h}_{11}, \label{23062018c}
\\
\superpotential{^{(1)(0)(2)}}=\superpotential{^{(1)(3)(2)}}=\superpotential{^{(2)(0)(1)}}=\superpotential{^{(2)(3)(1)}}=(\varepsilon/4c)\dot{h}_{12}. \label{23062018d}
\end{eqnarray}
Using Eq.~(\ref{22062018d}) to evaluate $\superpotential{^a^b^\lambda}$, we find that $\superpotential{^{(2)(2)t}}$, $\superpotential{^{(1)(0)x}}$, $\superpotential{^{(1)(3)x}}$, $\superpotential{^{(2)(2)z}}$, $-\superpotential{^{(1)(1)t}}$, $-\superpotential{^{(2)(0)y}}$, $-\superpotential{^{(2)(3)y}}$, and $-\superpotential{^{(1)(1)z}}$ are all equal to $(\varepsilon/4c)\dot{h}_{11}$, while $\superpotential{^{(2)(0)x}}$, $\superpotential{^{(2)(3)x}}$, $\superpotential{^{(1)(0)y}}$, $\superpotential{^{(1)(3)y}}$, $-\superpotential{^{(2)(1)t}}$, $-\superpotential{^{(1)(2)t}}$, $-\superpotential{^{(2)(1)z}}$,  and $-\superpotential{^{(1)(2)z}}$ are equal to $(\varepsilon/4c)\dot{h}_{12}$. Finally, we can calculate $t^{\lambda a}$ from these components and Eqs.~(\ref{22062018d})-(\ref{23062018d}). Doing so, we arrive at
\begin{equation}
t^{0(0)}=t^{3(0)}=t^{0(3)}=t^{3(3)}=\frac{c^2}{16\pi G}\left(\dot{h}_{11}^2+\dot{h}_{12}^2 \right),\label{23062018e}
\end{equation}
 where $\varepsilon$ has been absorbed into $h_{11}$ and $h_{12}$. This result is the same as that of GR \footnote{See, for example, Eq.~(1.136), page 36, of Ref.~\cite{MicheleMaggiore1}. }, except for averaging.
 
The energy density (\ref{23062018e}) is clearly positive, unlike the result of Ref.~\cite{ANDP:ANDP201200272}. This difference is a good example of the frame dependence of the gravitation energy.

\subsection{Exact plane GWs}\label{06072018b} 
To find the stress-energy of an exact plane GW, let us take the line element as
\begin{equation}
ds^2=dt^2-f(u)^2dx^2-g(u)^2dy^2-dz^2, \label{24062018a}
\end{equation}
where\footnote{The metric (\ref{24062018a}) will be a solution of the field equations only if $f$ and $g$ satisfy $\ddot{f}/f+\ddot{g}/g=0$ (see, e.g., page 280 of Ref.~\cite{Inverno}).} $\ddot{f}/f+\ddot{g}/g=0$  and $c=1$. This line element represents the propagation of a plane GW along the $z$ axis with a $+$ polarization. 

A convenient frame adapted to static observers for this spacetime is 
\begin{equation}
e_{a\mu}=diag(1,-f(u),-g(u),-1),\quad \Rightarrow\quad \e{_a^\mu}=diag(1,1/f(u),1/g(u),1). \label{24062018b}
\end{equation}
The Christoffel symbols for the spacetime (\ref{24062018a}) are 
\begin{eqnarray}
\tensor{\Gamma}{^x_t_x}=-\tensor{\Gamma}{^x_x_z}=\frac{\dot{f}}{f},\quad \tensor{\Gamma}{^t_x_x}=\tensor{\Gamma}{^z_x_x}=f\dot{f},\quad \tensor{\Gamma}{^y_t_y}=-\tensor{\Gamma}{^y_y_z}=\frac{\dot{g}}{g},\quad \tensor{\Gamma}{^t_y_y}=\tensor{\Gamma}{^z_y_y}=g\dot{g}. \label{11072018a}
\end{eqnarray}
Using Eqs.~(\ref{11072018a}) and (\ref{24062018b}) into Eq.~(\ref{23072018b}), one can verify that $\tensor{\phi}{_a^\lambda}=0$. This means that the frame (\ref{24062018b}) is a FFF (as a matter of fact, the wave does not even affect its motion).

The components of $\torsion{^a_b_c}$ and $\superpotential{^a^b^c}$ in this basis are
\begin{equation}
\torsion{^{(1)}_{(0)(1)}}=\torsion{^{(1)}_{(1)(3)}}=\dot{f}/f,\quad \torsion{^{(2)}_{(0)(2)}}=\torsion{^{(2)}_{(2)(3)}}=\dot{g}/g,  \label{24062018c}
\end{equation}
\begin{eqnarray}
\superpotential{^{(1)(0)(1)}}=\superpotential{^{(1)(3)(1)}}=\dot{g}/(2g),\quad \superpotential{^{(2)(0)(2)}}=\superpotential{^{(2)(3)(2)}}=\dot{f}/(2f),
\nonumber\\
\superpotential{^{(0)(0)(3)}}=\superpotential{^{(3)(0)(3)}}=\dot{f}/(2f)+\dot{g}/(2g). \label{24062018d}
\end{eqnarray}
In terms of $\superpotential{^a^b^\lambda}$, we have
\begin{eqnarray}
\superpotential{^{(1)(1)t}}=\superpotential{^{(1)(1)z}}=-\dot{g}/(2g),\quad \superpotential{^{(2)(2)t}}=\superpotential{^{(2)(2)z}}=-\dot{f}/(2f),
\\
\superpotential{^{(1)(0)x}}=\superpotential{^{(1)(3)x}}=\dot{g}/(2fg),\quad \superpotential{^{(2)(0)y}}=\superpotential{^{(2)(3)y}}=\dot{f}/(2fg),
\\
\superpotential{^{(0)(0)z}}=\superpotential{^{(3)(0)z}}=-\superpotential{^{(0)(3)t}}=-\superpotential{^{(3)(3)t}}=\dot{f}/(2f)+\dot{g}/(2g). \label{24062018e}
\end{eqnarray}
Finally, calculating $t^{\lambda a}$, we get
\begin{equation}
t^{0(0)}=t^{3(0)}=t^{0(3)}=t^{3(3)}=-\frac{c^2}{4\pi G}\frac{\dot{f}\dot{g}}{fg}. \label{24062018f}
\end{equation}
This expression holds for  any $f$ and $g$ that satisfies $\ddot{f}/f+\ddot{g}/g=0$. One cannot obtain such an expression in GR because a split between ripples and the background is necessary there, which is possible only when the conditions for the validity of the gravitational-wave formalism holds: the dimensionless amplitude of the wave must be much less than unity as well as the wavelength has to be much less than the background radius of curvature \cite{Gravitation}.

By taking $f(u)\approx (1-\varepsilon h_{11})^{1/2}$ and $g(u)\approx (1+\varepsilon h_{11})^{1/2}$, one can expand Eq.~(\ref{24062018f}) to second order and show that it reduces to  Eq.~(\ref{23062018e}) with $h_{12}=0$. This particular case corresponds the one that the authors in Ref.~\cite{0264-9381-20-21-008} used to obtain the energy flux.

The energy (\ref{24062018f}) can be clearly positive, as in the linearized example above. Again, this result is different from the one in Ref.~\cite{PhysRevD.78.047502} because the authors used a different set of observers\footnote{There are two other differences. In Ref.~\cite{PhysRevD.78.047502}, it is used a general $pp$-wave spacetime  and a different coordinate system. The relation between the coordinates used here and in Ref.~\cite{PhysRevD.78.047502}, for the $+$ polarization case, can be found in Ref.~\cite{Inverno}, page 281. }.

It is interesting to note that the energy of a GW as measured by an observer that is in free fall is not zero, unlike the case in the Schwarzschild spacetime \cite{ANDP:ANDP201200272}. One might think that this happens with the spacetime (\ref{24062018a}) because it is not static. However, as we will see in the next section, even in a static spacetime as the JNW one, the energy measured by freely falling observers are not necessarily zero.

\section{Gravitational Energy in JNW spacetime}
The action of the TEGR coupled with a massless scalar field $V$ can be written in the form
\begin{equation}
S=\int d^4x e\left(T+\mu V_{\alpha}V^{\alpha} \right), \label{02072018a}
\end{equation}
where $V_{\mu}=\partial_{\mu} V$. From this action we obtain the following field equations\footnote{From now on c=1.}:
\begin{equation}
\partial_{\nu}\left(e\superpotential{^a^\lambda^\nu} \right)=\frac{1}{4k}e\e{^a_\mu}\left(t^{\lambda\mu}+T^{\lambda\mu} \right), \label{02072018c}
\end{equation}
\begin{equation}
\rbox V=0, \label{02072018d}
\end{equation}
where $\rbox$ stands for the Riemannian d'Alembertian and 
\begin{equation}
T^{\lambda\mu}=2k\mu(V^{\lambda}V^{\mu}-\frac{1}{2}g^{\lambda\mu}V_{\alpha}V^{\alpha}). \label{02072018e}
\end{equation} 
These equations are equivalent to those of Einstein's equations minimally coupled with $V$. Their solution is known as Janis-Newman-Winicour-Wyman spacetime and can be expressed as
\begin{equation}
ds^2=W^Sdt^2-W^{-S}dr^2-r^2W^{1-S}d\Omega^2, \label{02072018f}
\end{equation}
\begin{equation}
V=-\frac{1}{r_0}\ln W,
\end{equation}
with $W=1-r_0/r$, $r_0=2\sqrt{M^2+\mu/2}$, $S=2M/r_0$, and $d\Omega^2=d\theta^2+\sin^2\theta d\phi^2$. For $0< S< 1$, we have a naked singularity at $r_0$, while $S=1$ ($\mu=0$) corresponds to the Schwarzschild case.
For more details, see Refs.~\cite{Wyman,PhysRevLett.20.878,doi:10.1142/S0217751X97002577,doi:10.1142/S0218271814500862,PhysRevD.83.087502}.

\subsection{The gravitational energy density in a FFF}\label{30072018a}

Using the definitions
\begin{equation}
W_n\equiv 1-\frac{(nS+1)r_0}{2r},\quad G\equiv \sqrt{1-W^S},\quad F\equiv W^{-S}G, \label{19072018a}
\end{equation}
we can write a FFF as
\begin{equation}
\e{_a^\mu}=\left(
\begin{array}{cccc}
W^{-S} & -G & 0 & 0
\\
-F\sin\theta\cos\phi & \sin\theta\cos\phi & \frac{\cos\theta\cos\phi}{rW^{(1-S)/2}} & -\frac{\sin\phi}{\sin\theta rW^{(1-S)/2}}
\\
-F\sin\theta\sin\phi & \sin\theta\sin\phi & \frac{\cos\theta\sin\phi}{rW^{(1-S)/2}} & \frac{\cos\phi}{\sin\theta rW^{(1-S)/2}}
\\
-F\cos\theta & \cos\theta & -\frac{\sin\theta}{rW^{(1-S)/2}} & 0
\end{array} \right). \label{19072018b}
\end{equation}
To check that $\phi_{ab}$ vanishes, we have to calculate  the Christoffel symbols for the metric (\ref{02072018f}). Doing so, we get:
\begin{equation}
\begin{array}{l}
\chr{^r_t_t}=Sr_0W^{2S-1}/(2r^2),\ \chr{^t_t_r}=-\chr{^r_r_r}=Sr_0W^{-1}/(2r^2),\ \chr{^r_\theta_\theta}=-rW_1,
\\
\chr{^r_\phi_\phi}=\sin^2\theta\chr{^r_\theta_\theta},\ \chr{^\theta_r_\theta}=\chr{^\phi_r_\phi}=\frac{W_1}{rW},\ \chr{^\theta_\phi_\phi}=-\sin^2\theta\chr{^\phi_\theta_\phi}=-\sin\theta\cos\theta.
\end{array}
\end{equation}
Then, using these symbols in Eq.~(\ref{23072018b}), one finds $\phi_{ab}=0$.

The torsion components in the frame (\ref{19072018b}) read:
\begin{equation}
\begin{array}{lll}
\torsion{^{(1)}_{(0)(1)}}&=&\Bigl\{\left[W_1-W_2\sin^2\theta\cos^2\phi\right]W^S
\\
&&+W_1\left(\sin^2\theta\cos^2\phi-1\right)\Bigr\}/(rWG),
\\ 
\torsion{^{(2)}_{(0)(2)}}&=&\Bigl\{\left[Sr_0/(2r)+W_2\left(1-\sin^2\theta\sin^2\phi\right)\right]W^S
\\
&&-W_1\left(1-\sin^2\theta\sin^2\phi\right)\Bigr\}/\left(rWG\right), 
\\ 
\torsion{^{(3)}_{(0)(3)}}&=&\left\{\left[W_1-W_2\cos^2\theta\right]W^S-W_1\sin^2\theta\right\}/(rWG),   
\\ 
\torsion{^{(2)}_{(0)(1)}}&=&\left\{\sin^2\theta\sin\phi\cos\phi \left[W_1-W_2W^S\right] \right\}/(rWG), 
\\ 
 \torsion{^{(1)}_{(1)(2)}}&=&\left\{\sin\theta\sin\phi \left[W^{(S+1)/2}-W_1\right] \right\}/(rW), 
\end{array} 
\end{equation}
\begin{equation}
\begin{array}{l}
 \torsion{^{(1)}_{(0)(3)}}=\torsion{^{(3)}_{(0)(1)}}= \torsion{^{(2)}_{(0)(1)}}\cos\theta/(\sin\theta\sin\phi),\ \torsion{^{(1)}_{(0)(2)}}=\torsion{^{(2)}_{(0)(1)}},
\\
\torsion{^{(2)}_{(0)(3)}}=\torsion{^{(3)}_{(0)(2)}}=\torsion{^{(2)}_{(0)(1)}}\cos\theta/(\sin\theta\cos\phi),\
\\
\torsion{^{(2)}_{(2)(3)}}=\torsion{^{(1)}_{(1)(3)}}=\torsion{^{(1)}_{(1)(2)}}\cos\theta/(\sin\theta\sin\phi),\ \torsion{^{(3)}_{(3)(2)}}=\torsion{^{(1)}_{(1)(2)}},
\\
 \torsion{^{(2)}_{(2)(1)}}=\torsion{^{(3)}_{(3)(1)}}=\torsion{^{(1)}_{(1)(2)}}\cos\phi/\sin\phi.
\end{array}\label{19072018h}
\end{equation}
It is clear that the torsion tensor vanishes for $r_0=0$ (in this case, $W=W_1=W_2=1$ ), which means that this tensor is measuring only the effects of gravity (the effects of gravity on a certain frame depends on how it is moving, though). Now, using these expressions in Eq.~(\ref{22072018d}), we discover that  the superpotential components are
\begin{equation}
\begin{array}{lll}
\superpotential{^{(0)(0)(1)}}&=&\left[W^{(S+1)/2}-W_1\right]\sin\theta\cos\phi/\left(rW\right),
\\
\superpotential{^{(1)(0)(1)}}&=&\Bigl\{\left(W_0+W_2\sin^2\theta\cos^2\phi \right)W^S
\\
&&-W_1\left(\sin^2\theta\cos^2\phi+1 \right) \Bigr\}/\left( 2rWG\right),
\\
\superpotential{^{(2)(0)(1)}}&=&\left(W_2W^S-W_1\right)\sin^2\theta\sin\phi\cos\phi/\left(2rWG\right),
\\
\superpotential{^{(2)(0)(2)}}&=&\Bigl\{\left(W_0+W_2\sin^2\theta\sin^2\phi \right)W^S
\\
&&-W_1\left(\sin^2\theta\sin^2\phi+1 \right) \Bigr\}/\left( 2rWG\right),
\\
\superpotential{^{(3)(0)(3)}}&=&\Bigl\{\left(W_0+W_2\cos^2\theta \right)W^S
\\
&&-W_1\left(\cos^2\theta+1 \right) \Bigr\}/\left( 2rWG\right),
\end{array}\label{21072018a}
\end{equation} 
\begin{equation}
\begin{array}{l}
\superpotential{^{(0)(0)(2)}}=2\superpotential{^{(1)(2)(1)}}=2\superpotential{^{(3)(2)(3)}}=\superpotential{^{(0)(0)(1)}}\sin\phi/\cos\phi,
\\ 
\superpotential{^{(0)(0)(3)}}=2\superpotential{^{(1)(3)(1)}}=2\superpotential{^{(2)(3)(2)}}=\superpotential{^{(0)(0)(1)}}\cos\theta/(\sin\theta\cos\phi),
\\
 \superpotential{^{(2)(1)(2)}}=\superpotential{^{(3)(1)(3)}}=(1/2)\superpotential{^{(0)(0)(1)}}, \ \superpotential{^{(1)(0)(2)}}=\superpotential{^{(2)(0)(1)}}
\\
\superpotential{^{(3)(0)(1)}}=\superpotential{^{(1)(0)(3)}}=\superpotential{^{(2)(0)(1)}}\cos\theta/(\sin\theta\sin\phi),
\\
\superpotential{^{(3)(0)(2)}}=\superpotential{^{(2)(0)(3)}}=\superpotential{^{(2)(0)(1)}}\cos\theta/(\sin\theta\cos\phi).
\end{array} \label{21072018b}
\end{equation}
Finally, we use the equations above in Eq.~(\ref{22072018c})  to find that $t^{0(0)}$ is not zero, but rather
given by
\begin{equation}
t^{0(0)}=\frac{k}{r^2W^2}\left\{W^{S+1}-\left[2W_1-W^{(S+1)/2}\right]^2-(S^2-1)r_0^2/(2r^2)\right\}. \label{21072018c}
\end{equation}
If we assume that Eq.~(\ref{21072018c}) represents the energy density of the gravitational field, the idea that a freely falling observer would not be able to measure the effects of gravity is not true, even for a static spacetime. It seems that the vanishing in the Schwarzschild case is just a coincidence. Nonetheless, this is not in contradiction with the principle of equivalence, since a FFF is not a local  inertial reference frame \cite{doi:10.1002/andp.201700175}.

It is easy to show that, in general, $t^{0(0)}+T^{0(0)}$ does not vanish either. Evaluating $\superpotential{^{(0)tr}}$ and using $e=r^2W^{1-S}\sin\theta$, we discover that the total energy of the spacetime within a sphere of radius $r$ is
\begin{equation}
E=16\pi krW^{-S}\left[-W_1+W^{(S+1)/2} \right], \label{21072018d}
\end{equation}
which vanishes for the Schwarzschild case, $S=1$; the same holds for Eq.~(\ref{21072018c}). It is worth noting that the total energy within the whole space is zero, that is, $E\to 0$ as $r\to\infty$.

\subsection{Static observers}\label{30072018b}
Let us now analyze the gravitational energy density as measured by a set of observers that are adapted to static observers at spacelike infinity:
\begin{equation}
\e{_a^\mu}=\left(
\begin{array}{cccc}
W^{-S/2} & 0 &0&0\\
0 &\sin\theta\cos\phi  W^{S/2} &\frac{\cos\theta\cos\phi}{r} W^{(S-1)/2} &-\frac{\sin\phi}{r\sin\theta}W^{(S-1)/2}\\
0&\sin\theta\sin\phi W^{S/2} &\frac{\cos\theta\sin\phi}{r}W^{(S-1)/2} &\frac{\cos\phi}{r\sin\theta}W^{(S-1)/2}\\
0&\cos\theta W^{S/2} &-\frac{\sin\theta}{r}W^{(S-1)/2}&0
\end{array}
\right).  \label{3072018a}
\end{equation}

From Eqs.~(\ref{22072018b}) and (\ref{3072018a}),  we find that 
\begin{equation}
\begin{array}{ll}
\torsion{^{(0)}_{(0)(1)}}=-\frac{Sr_0}{2r^2}W^{(S/2-1)}\sin\theta\cos\phi, &\torsion{^{(0)}_{(0)(2)}}=-\frac{Sr_0}{2r^2}W^{(S/2-1)}\sin\theta\sin\phi, 
\\
\torsion{^{(0)}_{(0)(3)}}=-\frac{Sr_0}{2r^2}W^{(S/2-1)}\cos\theta, & \torsion{^{(1)}_{(1)(2)}}=\torsion{^{(3)}_{(3)(2)}}=-f\sin\theta\sin\phi,
\\
\torsion{^{(2)}_{(1)(2)}}=\torsion{^{(3)}_{(1)(3)}}=f\sin\theta\cos\phi, & \torsion{^{(1)}_{(1)(3)}}=\torsion{^{(2)}_{(2)(3)}}=-f\cos\theta, \label{3072018b}
\end{array}
\end{equation}
where $f=W^{(S/2-1)}[W_1-W^{1/2}]/r$. Using (\ref{3072018b}) into (\ref{22072018d}) we get
\begin{equation}
\begin{array}{ll}
\superpotential{^{(0)(0)(1)}}=-f\sin\theta\cos\phi, & \superpotential{^{(0)(0)(2)}}=-f\sin\theta\sin\phi,
\\
\superpotential{^{(0)(0)(3)}}=-f\cos\theta, & \superpotential{^{(1)(1)(2)}}=\superpotential{^{(3)(3)(2)}}=h\sin\theta\sin\phi,
\\
\superpotential{^{(2)(1)(2)}}=\superpotential{^{(3)(1)(3)}}=-h\sin\theta\cos\phi, & \superpotential{^{(1)(1)(3)}}=\superpotential{^{(2)(2)(3)}}=h\cos\theta,
\end{array}, \label{3072018c} 
\end{equation}
where $h=W^{(S/2-1)}[W_0-W^{1/2}]/(2r)$.  Now we are able to calculate  $t^{\lambda a}$. Using the above expressions and Eq.~(\ref{22072018c}), we arrive at  
\begin{equation}
t^{0(0)}=-2k\frac{W^{(S/2-2)}}{r^2}\left(W_1-W^{1/2}\right)^2
\end{equation}
and $t^{0a}=0$. There are other components that do not vanish as well, but we omit them here. Notice that, unlike the linearized GW case, the component $t^{0(0)}$ is negative, at least for the cases where $S/2-2$ is not a fraction of odd numbers.

We can use Eqs.~(\ref{3072018a})-(\ref{3072018c}) to get
\begin{equation}
\begin{array}{lll}
\superpotential{^{(0)tr}}=-f, & \superpotential{^{(1)r\theta}}=-g\cos\theta\cos\phi, & \superpotential{^{(2)r\theta}}=-g\cos\theta\sin\phi,
\\
\superpotential{^{(3)r\theta}}=g\sin\theta, & \superpotential{^{(1)r\phi}}=g\sin\phi/\sin\theta, & \superpotential{^{(2)r\phi}}=-g\cos\phi/\sin\theta,
\end{array}\label{3072018d}
\end{equation} 
where $g=W^{[3(S-1)/2]}(W_0-W^{1/2})/(2r^2)$.  From (\ref{3072018a}) one also obtains $e=r^2W^{1-S}\sin\theta$.  

In order to calculate the  total energy of the spacetime (\ref{02072018f}) as measured by the static observers, we use Eqs.~(\ref{22072018g})-(\ref{23072018a}) and (\ref{3072018d}). It is clear from these equations that $P^{(j)}=0$, while
\begin{equation}
E=16\pi k r W^{-S/2}\left(W^{1/2}-W_1 \right). \label{4072018a}
\end{equation}
For $S=1$, we recover the Schwarzschild case in the TEGR, namely, $E=16\pi k r(1-W^{1/2})$. On the other hand, for $S\neq 0,1$, the total energy diverges as $r\to r_0=2M$ (Remember that, for $0<S<1$, $r_0$ corresponds to a naked singularity). We also recover the Schwarzschild case in the asymptotic limit, that is, $\displaystyle\lim_{r\to \infty}E=8\pi k Sr_0=16\pi kM=M $ (the rest energy of a particle).

The energy (\ref{4072018a}) is not the same as the one that is, in general, obtained in GR \cite{doi:10.1142/S0217751X97002577}. The method used in Ref.~\cite{doi:10.1142/S0217751X97002577} yields $E=M$ for an arbitrary $r$, which agrees with (\ref{4072018a}) only in the asymptotic limit.

\section{Conclusions}\label{30072018d}
We have seen that the EMTG of the TEGR for linearized GWs is identical to the one in GR, at least for the frame (\ref{22062018d}). As expected, the gravitational energy is positive. We have also seen that the EMTG yields an exact expression for the energy density of a linearly polarized plane GW that, in a frame with neither acceleration nor rotation (and also static), can be positive or negative. The expression turned out to be very simple and has no counterpart in GR.

The gravitational energy density of the JNW spacetime was obtained both in a static frame and in a FFF. Since this density did not vanish in the latter case, we conclude that, even in static spacetimes, observers in free fall can measure the gravitational energy. Nonetheless, this does not contradict the principle of equivalence. As pointed out in Ref.~\cite{doi:10.1002/andp.201700175}, this kind of frame is not really a local inertial reference frame. The total energy of the system within a sphere of radius $r$ was evaluated for both frames. For static observers, the total energy reduces to the value of the Schwarzschild case when $\mu=0$ (coupling constant), and also in the limit as $r\to\infty$. In turn, for the freely falling ones, the total energy of  the whole space turned out to be zero, as in the Schwarzschild case.

%

\end{document}